\documentclass[preprintnumbers,amsmath,amssymb,prd,floatfix,12pt,superscriptaddress,nofootinbib]{revtex4}
\usepackage{graphicx}
\usepackage{epsfig}
\usepackage{bm}
\usepackage{amsfonts}

\begin{document}
\title{Interacting entropy-corrected holographic Chaplygin gas model}
\author{M. Umar Farooq}
\email{mfarooq@camp.nust.edu.pk}
\affiliation{Center for Advanced Mathematics and Physics,\\
National University of Sciences and Technology, Rawalpindi, 46000, Pakistan}
\author{Muneer A. Rashid}
\affiliation{Center for Advanced Mathematics and Physics,\\
National University of Sciences and Technology, Rawalpindi, 46000, Pakistan}
\author{Mubasher Jamil}
\email{mjamil@camp.edu.pk}
\affiliation{Center for Advanced Mathematics and Physics,\\
National University of Sciences and Technology, Rawalpindi, 46000, Pakistan}

\begin{abstract}
\textbf{Abstract:} {\small Holographic dark energy (HDE), presents a
dynamical view of dark energy which is consistent with the observational
data and has a solid theoretical background. Its definition follows from the
entropy-area relation $S(A)$, where $S$ and $A$ are entropy and area
respectively. In the framework of loop quantum gravity, a modified
definition of HDE called ``entropy-corrected holographic dark energy''
(ECHDE) has been proposed recently to explain dark energy with the help of
quantum corrections to the entropy-area relation. Using this new definition,
we establish a correspondence between modified variable Chaplygin gas, new
modified Chaplygin gas and the viscous generalized Chaplygin gas with the
entropy corrected holographic dark energy and reconstruct the corresponding
scalar potentials which describe the dynamics of the scalar field. }
\end{abstract}

\maketitle

\newpage

\section{\textbf{Introduction}}

Numerous cosmological observations of type Ia supernova, cosmic microwave
background anisotropies measured with WMAP and large scale structure,
suggest that our universe is undergoing in an accelerated expansion possibly
due to the presence of dark energy which possesses negative pressure \cite%
{1}. In the astrophysics community, the nature of such mysterious
form of energy is still a matter of debate. In cosmology, several
candidates responsible for this expansion have been proposed namely,
Chaplygin gas, modified gravity, scalar tensor theory, tachyon and
braneworld model, to a name a few \cite{2}. Despite the fact that
the cosmological constant is the most obvious candidate that offers
a solution to the dark energy problem, yet it has several drawbacks
like fine-tuning and coincidence problems.

In order to soften the cosmic coincidence and fine tuning problems,
models of dark energy interacting with dark matter have been
proposed \cite{3}. It is generally asked that if the universe
evolves from the earlier quintessence ($\omega >-1$) to late
accelerating universe with phantom regime ($\omega <-1$), then why
$\omega =-1$ crossing occurs at the present time, where
 $\omega={p}/{\rho }$. The equation of state (EoS) parameter $\omega $
changes at different cosmic epoch which supports the evolving dark
energy \cite{4}. In the cosmological model explaining dark energy,
the ratio of energy densities of dark matter and dark energy
$\mathit{r}_{m}$ is of order unity, while observations suggest that
in pure dark energy model, this ratio must decrease. It may implies
the transfer of energy occurs between these two entities to keep a
subtle balance at the current time. The interaction between these
entities involves a coupling constant that determines the strength
of this interaction. Besides Einstein gravity, this study of
interaction can also be extended to $f(R)$, Brans-Dicke, braneworld,
Horava-Lifshitz and Gauss-Bonnet gravities \cite{5}.

In recent times, considerable interest has been stimulated to explain the
observed dark energy (dominant force in cosmos) with the help of holographic
dark energy model \cite{6}. According to holographic principle, the number
of degrees of freedom in a bounded physical system should be finite and has
relationship with the area of its boundary rather than with its volume \cite%
{7}. It is commonly believed that the holographic principle is a
fundamental principle of quantum gravity which is used to explain
the events involving high energy scale. It is motivated from an
observation that in quantum field theory, the ultra-violet cut-off
$\Lambda $ could be related to the infrared cut-off $L$ due to the
limit set by forming a black hole i.e. the quantum zero-point energy
of a system with size $L$ should not exceed the mass of a black hole
with the same size, i.e. $L^{3}\Lambda ^{3}\leq (M_{p}L)^{3/2}$
\cite{8}$.$ This last expression can be re-written as $L^{3}\rho
_{\Lambda }\leq LM_{p}^{2},$ where $\rho _{\Lambda }\backsim \Lambda
^{4}$ is the energy density corresponding to the zero point energy
and cut-off $\Lambda .$ Now the last inequality takes the form $\rho
_{\Lambda }\leq M_{p}^{2}L^{-2}$ or $\rho _{\Lambda
}=3n^{2}M_{p}^{2}L^{-2}.$ Here $3n^{2}$ is a constant and attached
for convenience. A sufficient literature is available on the study
of interaction between the holographic dark energy and the matter
\cite{9}. We are interested in studying the dynamics of entropy
corrected holographic dark energy (ECHDE) when it interacts with
some exotic type fluids. The entropy
corrected holographic dark energy is given by \cite{10}%
\begin{equation}
\rho _{\Lambda }=3n^{2}m_{p}^{2}L^{-2}+\gamma L^{-4}\ln
(m_{p}^{2}L^{2})+\beta L^{-4},
\end{equation}%
where the first term on right hand side is the usual holographic
dark energy while the other two appeared are mainly due to quantum
corrections in the loop quantum gravity (LQG). The correction term
play crucial role in early universe when L is small. When L gets
large, ECHDE reduces to HDE. The expression for the corrected
entropy of the black hole is given by $S=\frac{A}{4G}+\gamma \ln
(\frac{A}{4G})+\beta $ is
mainly arise due to the thermal equilibrium and quantum fluctuation \cite{11}%
. The parameters $n^{2},$ $\gamma $ and $\beta $ constant of order unity. If
we choose $\gamma $ and $\beta $ to be zero we arrive at the usual
holographic dark energy model. The parameter $n$ can be a function of time
\cite{12}, but in our discussion it is purely a constant quantity.

The plane of the paper is as follows: In section 2, the model of
Friedmann-Robertson-Walker (FRW) universe and the relevant equations are
presented. In section 3, we discuss a correspondence between ECHDE and
different Chaplygin gas forms including the modified variable Chaplygin gas
(MVCG) and new modified Chaplygin gas (NMCG) and the viscous generalized
Chaplygin gas (VGCG). In each case, we reconstruct the potentials and the
dynamics of the scalar field which describe the entropy corrected
holographic Chaplygin cosmology. The final section is devoted to the
conclusion.

\section{The model}

We start by assuming the background spacetime to be spatially
homogeneous and isotropic FRW spacetime given by
\begin{equation}
ds^{2}=-dt^{2}+a^{2}(t)\left[ \frac{dr^{2}}{1-kr^{2}}+r^{2}(d\theta
^{2}+\sin ^{2}\theta d\varphi ^{2})\right] .
\end{equation}%
Here $a(t)$ is the dimensionless scale factor which is an arbitrary function
of time and $k$ is defined to be the curvature parameter which has
dimensions of \emph{length}$^{-2}$ and its different values describe the
spatial geometry. For instance, for $k=-1,0,1,$ the above metric (2)
represents the spatially open, flat and closed FRW spacetimes respectively.
The Friedmann equation is given by%
\begin{equation}
H^{2}+\frac{k}{a^{2}}=\frac{1}{3m_{p}^{2}}[\rho _{\Lambda }+\rho _{m}],
\end{equation}%
where $m_{p}^{2}=(8\pi G)^{-1}$ is modified Planck mass. Here $H=\dot{a}/a$
is the Hubble parameter while $\rho _{\Lambda }$ and $\rho _{m}$ are the
energy densities of dark energy and matter respectively. In dimensionless
form, Eq. (3) can be written as%
\begin{equation}
1+\Omega _{k}=\Omega _{\Lambda }+\Omega _{m}.
\end{equation}%
The dimensionless density parameters corresponding to matter, dark energy
and curvature are%
\begin{equation}
\Omega _{m}=\frac{\rho _{m}}{\rho _{cr}}=\frac{\rho _{m}}{3H^{2}m_{p}^{2}},%
\text{ }\Omega _{\Lambda }=\frac{\rho _{\Lambda }}{\rho _{cr}}=\frac{\rho
_{\Lambda }}{3H^{2}m_{p}^{2}},\text{ }\Omega _{k}=\frac{k}{(aH)^{2}}.
\end{equation}%
Here $\rho _{cr}=3H^{2}m_{p}^{2}$ is the critical density. The energy
conservation equations for dark energy and dark matter are%
\begin{eqnarray}
\dot{\rho}_{\Lambda }+3H(\rho _{\Lambda }+p_{\Lambda }) &=&-Q, \\
\dot{\rho}_{m}+3H\rho _{m} &=&Q.
\end{eqnarray}%
Here overdot represents the differentiation with respect to cosmic
co-moving time $t$. Eq. (6) and (7) show that if there is an
interaction between dark energy and dark matter, the energy
conservation for dark energy and matter would not hold independently
but for the total interacting system. In explicit form, we have
$p_{\Lambda }=\omega _{\Lambda }\rho _{\Lambda }$ and $p_{m}=0.$
During the energy transfer, local energy conservation will not hold
in general but for the whole interacting system. Naturally if two
species are present in dominant form, it is obvious that they will
interact. If the quantity $Q$ is positive, it shows the transfer of
energy from dark matter to dark energy and vice versa in the case
when $Q$ is negative. The importance of interacting dark energy and
dark matter model also emerges since it is the best fit for the data
we obtain from the physical observations for instance SN Ia and
cosmic microwave background \cite{1}.

Defining the effective equation of state for dark energy and dark matter as
\cite{13}%
\begin{equation}
\omega _{\Lambda }^{\text{eff}}=\omega _{\Lambda }+\frac{\Gamma }{3H},\text{
}\omega _{m}^{\text{eff}}=-\frac{1}{r_{m}}\frac{\Gamma }{3H},
\end{equation}%
where the term $\Gamma =Q/\rho _{\Lambda }$ represents the decay
rate. After employing Eq. (8) in (6) and (7), we obtain the
following pair of
continuity equations%
\begin{eqnarray}
\dot{\rho}_{\Lambda }+3H(1+\omega _{\Lambda }^{\text{eff}})\rho _{\Lambda }
&=&0 \\
\dot{\rho}_{m}+3H(1+\omega _{m}^{\text{eff}})\rho _{m} &=&0.
\end{eqnarray}%
If we take $L$ as the Hubble scale $H^{-1}$ i.e. $L=H^{-1}$ at the present
epoch $H=H_{0}\backsim 10^{-33}eV,$ then the energy density $\rho _{\Lambda }
$ is comparable with the observed dark energy density $\backsim
10^{-10}eV^{4}.$ The second option for the infra-red cut-off is the particle
horizon. Hsu \cite{14} showed that under this scenario the resulting EoS
becomes zero and does not lead to an accelerated universe. So to get an
accelerated expansion of the observable universe, Li \cite{15} proposed that
the IR cut-off $L$ should be taken as the future event horizon and he
defined it as%
\begin{equation}
L=a(t)r(t),
\end{equation}%
here $r(t)$ is related to the future event horizon of the observable
universe. Using the FRW metric, we can obtain \cite{15}
\begin{equation}
L=a(t)\frac{\text{sinn}(\sqrt{|k|}y)}{\sqrt{|k|}},\ \ \ y=\frac{R_{h}}{a(t)},
\end{equation}%
where $R_{h}$ is the size of the future event horizon defined as
\begin{equation}
R_{h}=a(t)\int\limits_{t}^{\infty }\frac{dt^{\prime }}{a(t^{\prime })}%
=a(t)\int\limits_{0}^{r_{1}}\frac{dr}{\sqrt{1-kr^{2}}}.
\end{equation}%
The last integral has the explicit form as
\begin{equation}
\int\limits_{0}^{r_{1}}\frac{dr}{\sqrt{1-kr^{2}}}=\frac{1}{\sqrt{|k|}}\text{%
sinn}^{-1}(\sqrt{|k|}r_{1})=\begin{cases}\text{sin}^{-1}(r_1) , & \,
\,k=+1,\\ r_1, & \, \, k=0,\\ \text{sinh}^{-1}(r_1), & \, \,k=-1.\\
\end{cases}
\end{equation}%
Using the definition of $\rho _{\Lambda }$ and $\rho _{cr},$ we obtain the
following relation
\begin{equation}
HL=\sqrt{\frac{3n^{2}m_{p}^{2}+\gamma L^{-2}\ln (m_{p}^{2}L^{2})+\beta L^{-2}%
}{3m_{p}^{2}\Omega _{\Lambda }}}.
\end{equation}%
Differentiate eq. (12) with respect to time $t$\ and using Eq. (16), we have%
\begin{equation}
\dot{L}=\sqrt{\frac{3n^{2}m_{p}^{2}+\gamma L^{-2}\ln
(m_{p}^{2}L^{2})+\beta L^{-2}}{3m_{p}^{2}\Omega _{\Lambda
}}}-\text{cosn}(\sqrt{|k|}y),
\end{equation}%
where%
\begin{equation}
\text{cosn}(\sqrt{|k|}y)=\begin{cases} \text{cos}y , \, \,k=+1,\\ 1, \, \,
k=0,\\ \text{cosh}y, \, \,k=-1,\\ \end{cases}
\end{equation}%
After taking derivative of (1) with respect to $t,$ we get%
\begin{eqnarray}
\dot{\rho}_{\Lambda } &=&(2\gamma L^{-5}-4\gamma L^{-5}\ln
(m_{p}^{2}L^{2})-4\beta L^{-5}-6n^{2}m_{p}^{2}L^{-3})  \nonumber \\
&&\times \left[ \sqrt{\frac{3n^{2}m_{p}^{2}+\gamma L^{-2}\ln
(m_{p}^{2}L^{2})+\beta L^{-2}}{3m_{p}^{2}\Omega _{\Lambda }}}-\text{cosn}(%
\sqrt{|k|}y)\right] .
\end{eqnarray}%
Using Eq. (18) in (6), we arrive at
\begin{eqnarray}
w_{\Lambda } &=&-1-\Big(\frac{2\gamma L^{-2}-4\gamma L^{-2}\ln
(m_{p}^{2}L^{2})-4\beta L^{-2}-6n^{2}m_{p}^{2}}{3(3n^{2}m_{p}^{2}+\gamma
L^{-2}\ln (m_{p}^{2}L^{2})+\beta L^{-2})}\Big)  \nonumber \\
&&\times \left[ 1-\sqrt{\frac{3m_{p}^{2}\Omega _{\Lambda }}{%
3n^{2}m_{p}^{2}+\gamma L^{-2}\ln (m_{p}^{2}L^{2})+\beta L^{-2}}}\text{cosn}(%
\sqrt{|k|}y)\right] -\frac{b^{2}(1+\Omega _{k})}{\Omega _{\Lambda }}.
\end{eqnarray}%
Now making use of Eq. (19) in (8) yields%
\begin{eqnarray}
w_{\Lambda }^{\text{eff}} &=&-1-\Big(\frac{2\gamma L^{-2}-4\gamma L^{-2}\ln
(m_{p}^{2}L^{2})-4\beta L^{-2}-6n^{2}m_{p}^{2}}{3(3n^{2}m_{p}^{2}+\gamma
L^{-2}\ln (m_{p}^{2}L^{2})+\beta L^{-2})}\Big)  \nonumber \\
&&\times \left[ 1-\sqrt{\frac{3m_{p}^{2}\Omega _{\Lambda }}{%
3n^{2}m_{p}^{2}+\gamma L^{-2}\ln (m_{p}^{2}L^{2})+\beta L^{-2}}}\text{cosn}(%
\sqrt{|k|}y)\right] .
\end{eqnarray}

\section{Correspondence between ECHDE and Chaplygin gas variants}

Since there are several candidates of dark energy, so it is
essential to develop a correspondence and the relationships between
them. Kamenshchik et al \cite{16} studied a homogenous model based
on a single fluid obeying the EoS $p=-\frac{A_{0}}{\rho }$ called
the Chaplygin gas, where $p$ and $\rho $ represent the pressure and
energy density of the fluid and $A_{0}$ is some positive constant.
Possessing many physically interesting features, several authors
have used it to model the accelerated expansion of universe
\cite{17}. But it does not satisfactorily address the problems like
structure formation and cosmological perturbation power spectrum
\cite{18}. Subsequently, this equation was modified to the form
$p=-\frac{A_{0}}{\rho ^{\alpha }}$ called generalized Chaplygin gas
(GCG) to construct viable cosmological models. Two free parameters
involved in it: one is $A_{0}$ and the other $0\leq \alpha \leq 1.$
The GCG fluid behaves like dust for small size of the universe while
it acts as cosmological constant when universe gets sufficiently
large. The GCG equation has been further modified to $p=B\rho -\frac{A_{0}}{%
\rho ^{\alpha }}$ with $0\leq \alpha \leq 1,$ which is called
modified Chaplygin gas (MCG) \cite{19} and it involves \ three
parameters. An interesting feature connected with MCG equation of
state is that it shows radiations era in the early universe. At the
late time it behaves as cosmological constant which can be fitted to
a $\Lambda $CDM model. Late on, Guo and Jhang \cite{20} first
proposed a model $p=B\rho -\frac{A_{0}}{\rho ^{\alpha }}$ by taking
$A_{0}$ as a function of the cosmological scale factor $a(t)$ i.e.
$A_{0}=A_{0}(a(t)),$ which is known as modified variable Chaplygin
gas (MVCG) \cite{21}. This assumption seems to be reasonable since
$A_{0}(a(t))$ is related to scalar potential if we interpret
Chaplygin gas via Born-Infeld scalar field.

\subsection{Modified variable Chaplygin gas and ECHDE}

Suppose we have two species i.e. dark matter and dark energy. The later is
specified by the MVCG which is given by
\begin{equation}
p=B\rho _{\Lambda }-\frac{B_{0}a^{-\delta }}{\rho _{\Lambda
}^{\alpha }}.
\end{equation}%
The evolution of the energy density of MVCG is
\begin{equation}
\rho _{\Lambda }=\Big[\frac{3(\alpha +1)B_{0}}{[3(\alpha +1)(B+1)-\delta ]}%
\frac{1}{a^{\delta }}-\frac{C}{a^{3(\alpha +1)(B+1)}}\Big]^{\frac{1}{\alpha
+1}},
\end{equation}%
where $B_{0}$ and $C$ are some constants.

We now reconstruct expressions of the potential and the dynamics of the
scalar field in the presence of ECHDE. So for this, consider a time
dependent scalar field $\phi (t)$ with potential $V(\phi ),$ which are
directly related with the energy density and pressure of MVCG as%
\begin{eqnarray}
\rho _{\Lambda } &=&\frac{1}{2}\dot{\phi}^{2}+V(\phi ), \\
p_{\Lambda } &=&\frac{1}{2}\dot{\phi}^{2}-V(\phi ).
\end{eqnarray}%
Since the kinetic terms are positive hence it means that MVCG is of
quintessence type. Adding Eqs. (23) and (24), we get the kinetic term%
\begin{eqnarray}
\dot{\phi}^{2} &=&(1+B)\Big[\frac{3(\alpha +1)B_{0}}{[3(\alpha
+1)(B+1)-\delta ]}\frac{1}{a^{\delta }}-\frac{C}{a^{3(\alpha +1)(B+1)}}\Big]%
^{\frac{1}{\alpha +1}}  \nonumber \\
&&-\frac{B_{0}a^{-\delta }}{\Big[\frac{3(\alpha +1)B_{0}}{[3(\alpha
+1)(B+1)-\delta ]}\frac{1}{a^{\delta }}-\frac{C}{a^{3(\alpha +1)(B+1)}}\Big]%
^{\frac{\alpha }{\alpha +1}}}.
\end{eqnarray}%
Subtraction of Eqs. (23) and (24) yields the potential term%
\begin{eqnarray}
2V(\phi ) &=&(1-B)\Big[\frac{3(\alpha +1)B_{0}}{[3(\alpha +1)(B+1)-\delta ]}%
\frac{1}{a^{n}}-\frac{C}{a^{3(\alpha +1)(B+1)}}\Big]^{\frac{1}{\alpha +1}}
\nonumber \\
&&+\frac{B_{0}a^{-\delta }}{\Big[\frac{3(\alpha +1)B_{0}}{[3(\alpha
+1)(B+1)-\delta ]}\frac{1}{a^{\delta }}-\frac{C}{a^{3(\alpha +1)(B+1)}}\Big]%
^{\frac{\alpha }{\alpha +1}}}.
\end{eqnarray}%
To see the correspondence between the ECHDE and MVCG energy density, we use
Eqs. (1) and (22) to get
\begin{equation}
C=a^{3(\alpha +1)(B+1)}\Big[\frac{3(\alpha +1)B_{0}}{[3(\alpha
+1)(B+1)-\delta ]}\frac{1}{a^{\delta }}-(3n^{2}m_{p}^{2}L^{-2}+\gamma
L^{-4}\ln (m_{p}^{2}L^{2})+\beta L^{-4})^{\alpha +1}\Big].
\end{equation}%
Writing Eq. (21) in an alternate form%
\begin{equation}
B_{0}=a^{\delta }(B-w_{\Lambda })\rho _{\Lambda }^{\alpha +1}.
\end{equation}%
Therefore in view of Eq. (19), the above equation (28) gives the value of
the parameter $B_{0}$ as%
\begin{eqnarray}
B_{0} &=&a^{\delta }(3n^{2}m_{p}^{2}L^{-2}+\gamma L^{-4}\ln
(m_{p}^{2}L^{2})+\beta L^{-4})^{\alpha +1}\Big[1+B  \nonumber \\
&&+\frac{2\gamma L^{-2}-4\gamma L^{-2}\ln (m_{p}^{2}L^{2})-4\beta
L^{-2}-6n^{2}m_{p}^{2}}{3(3n^{2}m_{p}^{2}+\gamma L^{-2}\ln
(m_{p}^{2}L^{2})+\beta L^{-2})}  \nonumber \\
&&\times \left( 1-\sqrt{\frac{3m_{p}^{2}\Omega _{\Lambda }}{%
3n^{2}m_{p}^{2}+\gamma L^{-2}\ln (m_{p}^{2}L^{2})+\beta L^{-2}}}\text{cosn}(%
\sqrt{|k|}y)\right) +\frac{b^{2}(1+\Omega _{k})}{\Omega _{\Lambda }}\Big].
\end{eqnarray}%
Now to determine the other parameter $C$ substitute the value of $B_{0}$ in
Eq. (27), we have%
\begin{eqnarray}
C &=&a^{3(\alpha +1)(A+1)}(3n^{2}m_{p}^{2}L^{-2}+\gamma L^{-4}\ln
(m_{p}^{2}L^{2})+\beta L^{-4})^{\alpha +1}\Big(\frac{3(\alpha +1)}{[3(\alpha
+1)(B+1)-\delta ]}  \nonumber \\
&&\times (B+1+\frac{2\gamma L^{-2}-4\gamma L^{-2}\ln (m_{p}^{2}L^{2})-4\beta
L^{-2}-6n^{2}m_{p}^{2}}{3(3n^{2}m_{p}^{2}+\gamma L^{-2}\ln
(m_{p}^{2}L^{2})+\beta L^{-2})})  \nonumber \\
&&\times \left[ 1-\sqrt{\frac{3m_{p}^{2}\Omega _{\Lambda }}{%
3n^{2}m_{p}^{2}+\gamma L^{-2}\ln (m_{p}^{2}L^{2})+\beta L^{-2}}}\text{cosn}(%
\sqrt{|k|}y)\right] -1\Big)
\end{eqnarray}%
Now we can re-write the kinetic energy and scalar potential terms as
\begin{eqnarray}
\dot{\phi}^{2} &=&-(3n^{2}m_{p}^{2}L^{-2}+\gamma L^{-4}\ln
(m_{p}^{2}L^{2})+\beta L^{-4})  \nonumber \\
&&\Big(\frac{2\gamma L^{-2}-4\gamma L^{-2}\ln (m_{p}^{2}L^{2})-4\beta
L^{-2}-6n^{2}m_{p}^{2}}{3(3n^{2}m_{p}^{2}+\gamma L^{-2}\ln
(m_{p}^{2}L^{2})+\beta L^{-2})}  \nonumber \\
&&\times \left[ 1-\sqrt{\frac{3m_{p}^{2}\Omega _{\Lambda }}{%
3n^{2}m_{p}^{2}+\gamma L^{-2}\ln (m_{p}^{2}L^{2})+\beta L^{-2}}}\text{cosn}(%
\sqrt{|k|}y)\right] +\frac{b^{2}(1+\Omega _{k})}{\Omega _{\Lambda }}\Big),
\end{eqnarray}%
and%
\begin{eqnarray}
2V(\phi ) &=&(3n^{2}m_{p}^{2}L^{-2}+\gamma L^{-4}\ln (m_{p}^{2}L^{2})+\beta
L^{-4})  \nonumber \\
&&\times \Big(2+\frac{2\gamma L^{-2}-4\gamma L^{-2}\ln
(m_{p}^{2}L^{2})-4\beta L^{-2}-6n^{2}m_{p}^{2}}{3(3n^{2}m_{p}^{2}+\gamma
L^{-2}\ln (m_{p}^{2}L^{2})+\beta L^{-2})}  \nonumber \\
&&\times \left[ 1-\sqrt{\frac{3m_{p}^{2}\Omega _{\Lambda }}{%
3n^{2}m_{p}^{2}+\gamma L^{-2}\ln (m_{p}^{2}L^{2})+\beta L^{-2}}}\text{cosn}(%
\sqrt{|k|}y)\right] +\frac{b^{2}(1+\Omega _{k})}{\Omega _{\Lambda }}\Big)
\end{eqnarray}%
Using the relation $x=\ln a,$ we get $\dot{\phi}=\phi ^{\prime }H,$ where $%
^{\prime }$ represents the derivative with respect to e-folding time
parameter $\ln a.$ After putting the value of $\dot{\phi}$ and applying
integration, we obtain%
\begin{eqnarray}
\phi (a)-\phi (a_{0}) &=&\frac{1}{H}\int_{0}^{\ln
a}[-(3n^{2}m_{p}^{2}L^{-2}+\gamma L^{-4}\ln (m_{p}^{2}L^{2})+\beta L^{-4})
\nonumber \\
&&\times \Big(\frac{2\gamma L^{-2}-4\gamma L^{-2}\ln (m_{p}^{2}L^{2})-4\beta
L^{-2}-6n^{2}m_{p}^{2}}{3(3n^{2}m_{p}^{2}+\gamma L^{-2}\ln
(m_{p}^{2}L^{2})+\beta L^{-2})}  \nonumber \\
&&\times \left[ 1-\sqrt{\frac{3m_{p}^{2}\Omega _{\Lambda }}{%
3n^{2}m_{p}^{2}+\gamma L^{-2}\ln (m_{p}^{2}L^{2})+\beta L^{-2}}}\text{cosn}(%
\sqrt{|k|}y)\right] \Big)+\frac{b^{2}(1+\Omega _{k})}{\Omega _{\Lambda }}%
]^{1/2}d\ln a,  \nonumber \\
&&
\end{eqnarray}%
where $a_{0}$ is the present value of the scale factor.

\subsection{New modified Chaplygin gas model and ECHDE}

The model that represents the dark energy is now the new modified Chaplygin
gas (NMCG) given by \cite{22}%
\begin{equation}
p_{\Lambda }=B\rho _{\Lambda }-\frac{K(a)}{\rho _{\Lambda }^{\alpha }},\text{
\ \ where }B>0\text{ and }0\leq \alpha \leq 1.
\end{equation}%
Here $K(a)$ is a function of scale factor of the universe. Taking $K(a)$ in
the form $K(a)=-\omega _{\Lambda }A_{1}a^{-3(w_{\Lambda }+1)(\alpha +1)}$ as
introduced by the authors in \cite{23}, we get%
\begin{equation}
p_{\Lambda }=B\rho _{\Lambda }+\frac{w_{\Lambda }A_{1}}{\rho _{\Lambda
}^{\alpha }}a^{-3(w_{\Lambda }+1)(\alpha +1)}.
\end{equation}%
The energy density of the NMCG can be expressed as%
\begin{equation}
\rho _{\Lambda }=\Big[\frac{w_{\Lambda }}{w_{\Lambda }-B}A_{1}a^{-3(w_{%
\Lambda }+1)(\alpha +1)}+B_{1}a^{-3(B+1)(\alpha +1)}\Big]^{\frac{1}{\alpha +1%
}},
\end{equation}%
where $B_{1}$ is a constant of integration. Following the previous
section we establish the correspondence between the ECHDE and NMCG
energy density.
Comparing Eqs. (36) and (1), we get%
\begin{eqnarray}
B_{1} &=&a^{3(B+1)(\alpha +1)}\Big((3n^{2}m_{p}^{2}L^{-2}+\gamma L^{-4}\ln
(m_{p}^{2}L^{2})+\beta L^{-4})^{\alpha +1}  \nonumber \\
&&-\frac{1+\frac{2\gamma L^{-2}-4\gamma L^{-2}\ln (m_{p}^{2}L^{2})-4\beta
L^{-2}-6n^{2}m_{p}^{2}}{3(3n^{2}m_{p}^{2}+\gamma L^{-2}\ln
(m_{p}^{2}L^{2})+\beta L^{-2})}\times \left[ 1-\sqrt{\frac{3m_{p}^{2}\Omega
_{\Lambda }}{3n^{2}m_{p}^{2}+\gamma L^{-2}\ln (m_{p}^{2}L^{2})+\beta L^{-2}}}%
\text{cosn}(\sqrt{|k|}y)\right] }{1+B+\frac{2\gamma L^{-2}-4\gamma L^{-2}\ln
(m_{p}^{2}L^{2})-4\beta L^{-2}-6n^{2}m_{p}^{2}}{3(3n^{2}m_{p}^{2}+\gamma
L^{-2}\ln (m_{p}^{2}L^{2})+\beta L^{-2})}\times \left[ 1-\sqrt{\frac{%
3m_{p}^{2}\Omega _{\Lambda }}{3n^{2}m_{p}^{2}+\gamma L^{-2}\ln
(m_{p}^{2}L^{2})+\beta L^{-2}}}\text{cosn}(\sqrt{|k|}y)\right] }  \nonumber
\\
&&\times \Big(A_{1}a^{3(\alpha +1)\frac{2\gamma L^{-2}-4\gamma L^{-2}\ln
(m_{p}^{2}L^{2})-4\beta L^{-2}-6n^{2}m_{p}^{2}}{3(3n^{2}m_{p}^{2}+\gamma
L^{-2}\ln (m_{p}^{2}L^{2})+\beta L^{-2})}\times \left[ 1-\sqrt{\frac{%
3m_{p}^{2}\Omega _{\Lambda }}{3n^{2}m_{p}^{2}+\gamma L^{-2}\ln
(m_{p}^{2}L^{2})+\beta L^{-2}}}\text{cosn}(\sqrt{|k|}y)\right] }\Big)\Big).
\end{eqnarray}%
With the use of $p_{\Lambda }=\omega _{\Lambda }\rho _{\Lambda },$ Eq. (35)
gives an expression for $A_{1}$ i.e.%
\begin{eqnarray}
A_{1} &=&(3n^{2}m_{p}^{2}L^{-2}+\gamma L^{-4}\ln (m_{p}^{2}L^{2})+\beta
L^{-4})^{\alpha +1}  \nonumber \\
&&\times a^{-3(\alpha +1)\frac{2\gamma L^{-2}-4\gamma L^{-2}\ln
(m_{p}^{2}L^{2})-4\beta L^{-2}-6n^{2}m_{p}^{2}}{3(3n^{2}m_{p}^{2}+\gamma
L^{-2}\ln (m_{p}^{2}L^{2})+\beta L^{-2})}\times \Big(1-\sqrt{\frac{%
3m_{p}^{2}\Omega _{\Lambda }}{3n^{2}m_{p}^{2}+\gamma L^{-2}\ln
(m_{p}^{2}L^{2})+\beta L^{-2}}}\text{cosn}(\sqrt{|k|}y)\Big)}  \nonumber \\
&&\times \left[ \frac{B+1+\frac{2\gamma L^{-2}-4\gamma L^{-2}\ln
(m_{p}^{2}L^{2})-4\beta L^{-2}-6n^{2}m_{p}^{2}}{3(3n^{2}m_{p}^{2}+\gamma
L^{-2}\ln (m_{p}^{2}L^{2})+\beta L^{-2})}\times \left[ 1-\sqrt{\frac{%
3m_{p}^{2}\Omega _{\Lambda }}{3n^{2}m_{p}^{2}+\gamma L^{-2}\ln
(m_{p}^{2}L^{2})+\beta L^{-2}}}\text{cosn}(\sqrt{|k|}y)\right] }{1+\frac{%
2\gamma L^{-2}-4\gamma L^{-2}\ln (m_{p}^{2}L^{2})-4\beta
L^{-2}-6n^{2}m_{p}^{2}}{3(3n^{2}m_{p}^{2}+\gamma L^{-2}\ln
(m_{p}^{2}L^{2})+\beta L^{-2})}\times \left[ 1-\sqrt{\frac{3m_{p}^{2}\Omega
_{\Lambda }}{3n^{2}m_{p}^{2}+\gamma L^{-2}\ln (m_{p}^{2}L^{2})+\beta L^{-2}}}%
\text{cosn}(\sqrt{|k|}y)\right] }\right] .  \nonumber \\
&&
\end{eqnarray}%
So we have found the expressions for the constants $A_{1}$ and $B_{1}.$
Employing Eqs. (23) and (24), the kinetic and potential terms are found to be%
\begin{eqnarray}
\dot{\phi}^{2} &=&(B+1)\Big[\frac{w_{\Lambda }}{w_{\Lambda }-B}%
A_{1}a^{-3(w_{\Lambda }+1)(\alpha +1)}+B_{1}a^{-3(B+1)(\alpha +1)}\Big]^{%
\frac{1}{\alpha +1}}  \nonumber \\
&&+\frac{A_{1}w_{\Lambda }a^{-3(w_{\Lambda }+1)(\alpha +1)}}{[\frac{%
w_{\Lambda }}{w_{\Lambda }-B}A_{1}a^{-3(w_{\Lambda }+1)(\alpha
+1)}+B_{1}a^{-3(B+1)(\alpha +1)}]^{\frac{\alpha }{\alpha +1}}},
\end{eqnarray}%
and
\begin{eqnarray}
2V(\phi ) &=&(1-B)[\frac{w_{\Lambda }}{w_{\Lambda }-B}A_{1}a^{-3(w_{\Lambda
}+1)(\alpha +1)}+B_{1}a^{-3(B+1)(\alpha +1)}]^{\frac{1}{\alpha +1}}
\nonumber \\
&&-\frac{A_{1}w_{\Lambda }a^{-3(w_{\Lambda }+1)(\alpha +1)}}{[\frac{%
w_{\Lambda }}{w_{\Lambda }-B}A_{1}a^{-3(w_{\Lambda }+1)(\alpha
+1)}+B_{1}a^{-3(B+1)(\alpha +1)}]^{\frac{\alpha }{\alpha +1}}},
\end{eqnarray}%
where $A_{1}$ and $B_{1}$ are given in (38) and (37) respectively.
In all Eqs. (35), (36), (39) and (40) the value of $ w_{\Lambda }$
is given by Eq. (19). From Eqs. (23) and
(24), the kinetic energy term is re-written to be%
\begin{eqnarray}
\dot{\phi}^{2} &=&-(3n^{2}m_{p}^{2}L^{-2}+\gamma L^{-4}\ln
(m_{p}^{2}L^{2})+\beta L^{-4})  \nonumber \\
&&\times \Big(\frac{2\gamma L^{-2}-4\gamma L^{-2}\ln (m_{p}^{2}L^{2})-4\beta
L^{-2}-6n^{2}m_{p}^{2}}{3(3n^{2}m_{p}^{2}+\gamma L^{-2}\ln
(m_{p}^{2}L^{2})+\beta L^{-2})}  \nonumber \\
&&\times \left[ 1-\sqrt{\frac{3m_{p}^{2}\Omega _{\Lambda }}{%
3n^{2}m_{p}^{2}+\gamma L^{-2}\ln (m_{p}^{2}L^{2})+\beta L^{-2}}}\text{cosn}(%
\sqrt{|k|}y)\right] +\frac{b^{2}(1+\Omega _{k})}{\Omega _{\Lambda }}\Big),
\end{eqnarray}%
while the potential energy term has the form%
\begin{eqnarray}
2V(\phi ) &=&(3n^{2}m_{p}^{2}L^{-2}+\gamma L^{-4}\ln (m_{p}^{2}L^{2})+\beta
L^{-4})  \nonumber \\
&&\times \Big(2+\frac{2\gamma L^{-2}-4\gamma L^{-2}\ln
(m_{p}^{2}L^{2})-4\beta L^{-2}-6n^{2}m_{p}^{2}}{3(3n^{2}m_{p}^{2}+\gamma
L^{-2}\ln (m_{p}^{2}L^{2})+\beta L^{-2})}  \nonumber \\
&&\times \left[ 1-\sqrt{\frac{3m_{p}^{2}\Omega _{\Lambda }}{%
3n^{2}m_{p}^{2}+\gamma L^{-2}\ln (m_{p}^{2}L^{2})+\beta L^{-2}}}\text{cosn}(%
\sqrt{|k|}y)\right] +\frac{b^{2}(1+\Omega _{k})}{\Omega _{\Lambda }}\Big).
\nonumber \\
&&
\end{eqnarray}%
Following the same steps as done for the MVCG, the kinetic term is easily
transformable to the following form%
\begin{eqnarray}
\phi (a)-\phi (a_{0}) &=&\frac{1}{H}\int_{0}^{\ln
a}[-(3n^{2}m_{p}^{2}L^{-2}+\gamma L^{-4}\ln (m_{p}^{2}L^{2})+\beta L^{-4})
\nonumber \\
&&\times \frac{2\gamma L^{-2}-4\gamma L^{-2}\ln (m_{p}^{2}L^{2})-4\beta
L^{-2}-6n^{2}m_{p}^{2}}{3(3n^{2}m_{p}^{2}+\gamma L^{-2}\ln
(m_{p}^{2}L^{2})+\beta L^{-2})}  \nonumber \\
&&\times \left[ 1-\sqrt{\frac{3m_{p}^{2}\Omega _{\Lambda }}{%
3n^{2}m_{p}^{2}+\gamma L^{-2}\ln (m_{p}^{2}L^{2})+\beta L^{-2}}}\text{cosn}(%
\sqrt{|k|}y)\right] +\frac{b^{2}(1+\Omega _{k})}{\Omega _{\Lambda }}%
]^{1/2}d\ln a.  \nonumber \\
&&
\end{eqnarray}

\subsection{Viscous generalized Chaplygin gas model and ECHDE}

If we assume that the dark energy as non-viscous as taken in section A and
B, it must results in the occurrence of a cosmic singularity (Big Rip) in
the far future. This singularity can be alleviated by introducing the
quantum corrections due to the conformal anomaly while the other option is
to consider the bulk viscosity $\xi $ of the cosmic fluid \cite{24}. The
theory of bulk viscosity was initially investigated by Eckart and later on
pursued by Landau and Lifshitz \cite{25}. The important feature of dark
energy with bulk viscosity is that it shows the accelerated expansion of
phantom type in the later epoch and softens the coincidence problem, age
problem and phantom crossing \cite{26}. The effective pressure containing
the isotropic pressure and viscous stress is given by the equation of state%
\begin{equation}
p_{\text{eff}}=p_{\Lambda }+\Pi ,
\end{equation}%
where $p_{\Lambda }=\frac{\chi }{\rho _{\Lambda }^{\alpha }},$ $\chi >0.$
Notice that first term on the right hand side mimics the GCG and the
parameter $\alpha $ varies as $0<\alpha \leq 1.$ If $\alpha =1$ it
represents the Chaplygin gas model. On the other hand if $\alpha <0$, it
corresponds to a polytropic gas. The bulk viscous fluid is represented by $%
\Pi =-\xi (\rho _{\Lambda })u_{;\mu }^{\mu }$ where $u^{\mu }$ is the
four-velocity vector of the viscous fluid and $\xi >0$ to get the positive
entropy production in conformity with second law of thermodynamics \cite{27}%
. We choose $\xi (\rho _{\Lambda })=\nu \rho _{\Lambda }^{1/2}$,
with $\nu $ as constant. The energy conservation equation yields the
energy density of
VGCG as%
\begin{equation}
\rho _{\Lambda }=[\frac{Da^{-3(\alpha +1)(1-\nu \gamma _{1})}-\chi }{(1-\nu
\gamma _{1})}]^{\frac{1}{\alpha +1}},
\end{equation}
\cite{28}. Here $\gamma _{1}=m_{p}^{-1}\sqrt{1-r_{m}},$ where $r_{m}=\frac{%
\rho _{m}}{\rho _{\Lambda }}=\frac{\Omega _{m}}{\Omega _{\Lambda }}$ and $D$
some constant of integration. The effective equation of state (45) says that
the universe is expanding in accelerated manner. It has been shown that this
phenomenon of universe can be studied with the help of dynamical evolving
scalar fields usually called inflation. So it was introduced to construct
models dealing with minimally coupled scalar field. Hence we construct the
dynamical scalar field $\phi $ with potential $V(\phi ),$ related to the
energy density and pressure of viscous dark energy model as%
\begin{eqnarray}
\rho _{\Lambda } &=&[\frac{Da^{-3(\alpha +1)(1-\nu \gamma _{1})}-\chi }{%
(1-\nu \gamma _{1})}]^{\frac{1}{\alpha +1}},  \nonumber \\
p_{\Lambda } &=&\chi \lbrack \frac{(1-\nu \gamma _{1})}{Da^{-3(\alpha
+1)(1-\nu \gamma _{1})}-\chi }]^{\frac{\alpha }{\alpha +1}}-3\nu H[\frac{%
Da^{-3(\alpha +1)(1-\nu \gamma _{1})}-\chi }{(1-\nu \gamma _{1})}]^{1/2}.
\end{eqnarray}%
To determine the expressions for the unknown $D$ and $\chi $ we take the
following steps. The effective equation of state for the interacting VGCG is%
\begin{equation}
w_{\Lambda }^{\text{eff}}=\frac{\chi }{\rho _{\Lambda }^{\alpha +1}}%
-3\upsilon H\rho _{\Lambda }^{-\frac{1}{2}}+\frac{b^{2}(1+\Omega _{k})}{%
\Omega _{\Lambda }}.
\end{equation}%
After substituting the value of $w_{\Lambda }^{\text{eff}}$ in (47) and
making simplification, we obtain%
\begin{eqnarray}
\chi  &=&(3n^{2}m_{p}^{2}L^{-2}+\gamma L^{-4}\ln (m_{p}^{2}L^{2})+\beta
L^{-4})^{\alpha +1}  \nonumber \\
&&\times \Big(-1-\frac{2\gamma L^{-2}-4\gamma L^{-2}\ln
(m_{p}^{2}L^{2})-4\beta L^{-2}-6n^{2}m_{p}^{2}}{3(3n^{2}m_{p}^{2}+\gamma
L^{-2}\ln (m_{p}^{2}L^{2})+\beta L^{-2})}  \nonumber \\
&&\times \left[ 1-\sqrt{\frac{3m_{p}^{2}\Omega _{\Lambda }}{%
3n^{2}m_{p}^{2}+\gamma L^{-2}\ln (m_{p}^{2}L^{2})+\beta L^{-2}}}\text{cosn}(%
\sqrt{|k|}y)\right]   \nonumber \\
&&+3\upsilon H(3n^{2}m_{p}^{2}L^{-2}+\gamma L^{-4}\ln (m_{p}^{2}L^{2})+\beta
L^{-4})^{-1/2}-\frac{b^{2}(1+\Omega _{k})}{\Omega _{\Lambda }}\Big).
\nonumber \\
&&
\end{eqnarray}%
Inserting the value of $\chi $ in (45), we get%
\begin{eqnarray}
D &=&a^{3(\alpha +1)(1-\nu \gamma _{1})}(3n^{2}m_{p}^{2}L^{-2}+\gamma
L^{-4}\ln (m_{p}^{2}L^{2})+\beta L^{-4})^{\alpha +1}  \nonumber \\
&&\times \lbrack -\nu \gamma _{1}-\frac{2\gamma L^{-2}-4\gamma L^{-2}\ln
(m_{p}^{2}L^{2})-4\beta L^{-2}-6n^{2}m_{p}^{2}}{3(3n^{2}m_{p}^{2}+\gamma
L^{-2}\ln (m_{p}^{2}L^{2})+\beta L^{-2})}  \nonumber \\
&&\times \left[ 1-\sqrt{\frac{3m_{p}^{2}\Omega _{\Lambda }}{%
3n^{2}m_{p}^{2}+\gamma L^{-2}\ln (m_{p}^{2}L^{2})+\beta L^{-2}}}\text{cosn}(%
\sqrt{|k|}y)\right]   \nonumber \\
&&+3\upsilon H(3n^{2}m_{p}^{2}L^{-2}+\gamma L^{-4}\ln (m_{p}^{2}L^{2})+\beta
L^{-4})^{-\frac{1}{2}}-\frac{b^{2}(1+\Omega _{k})}{\Omega _{\Lambda }}].
\nonumber \\
&&
\end{eqnarray}%
Now we can re-write the scalar potential and kinetic energy terms as
following%
\begin{eqnarray}
2V(\phi ) &=&(3n^{2}m_{p}^{2}L^{-2}+\gamma L^{-4}\ln (m_{p}^{2}L^{2})+\beta
L^{-4})  \nonumber \\
&&\times (2+\frac{2\gamma L^{-2}-4\gamma L^{-2}\ln (m_{p}^{2}L^{2})-4\beta
L^{-2}-6n^{2}m_{p}^{2}}{3(3n^{2}m_{p}^{2}+\gamma L^{-2}\ln
(m_{p}^{2}L^{2})+\beta L^{-2})}  \nonumber \\
&&\times \left[ 1-\sqrt{\frac{3m_{p}^{2}\Omega _{\Lambda }}{%
3n^{2}m_{p}^{2}+\gamma L^{-2}\ln (m_{p}^{2}L^{2})+\beta L^{-2}}}\text{cosn}(%
\sqrt{|k|}y)\right] +\frac{b^{2}(1+\Omega _{k})}{\Omega _{\Lambda }}).
\nonumber \\
&&
\end{eqnarray}%
and%
\begin{eqnarray}
\dot{\phi}^{2} &=&-(3n^{2}m_{p}^{2}L^{-2}+\gamma L^{-4}\ln
(m_{p}^{2}L^{2})+\beta L^{-4})  \nonumber \\
&&\times \Big(\frac{2\gamma L^{-2}-4\gamma L^{-2}\ln (m_{p}^{2}L^{2})-4\beta
L^{-2}-6n^{2}m_{p}^{2}}{3(3n^{2}m_{p}^{2}+\gamma L^{-2}\ln
(m_{p}^{2}L^{2})+\beta L^{-2})}  \nonumber \\
&&\times \left[ 1-\sqrt{\frac{3m_{p}^{2}\Omega _{\Lambda }}{%
3n^{2}m_{p}^{2}+\gamma L^{-2}\ln (m_{p}^{2}L^{2})+\beta L^{-2}}}\text{cosn}(%
\sqrt{|k|}y)\right] +\frac{b^{2}(1+\Omega _{k})}{\Omega _{\Lambda }}\Big).
\end{eqnarray}%
Again using the equation $\dot{\phi}=\phi ^{\prime }H,$ we can write%
\begin{eqnarray}
\phi (a)-\phi (a_{0}) &=&\frac{1}{H}\int_{0}^{\ln
a}[-(3n^{2}m_{p}^{2}L^{-2}+\gamma L^{-4}\ln (m_{p}^{2}L^{2})+\beta L^{-4})
\nonumber \\
&&\times (\frac{2\gamma L^{-2}-4\gamma L^{-2}\ln (m_{p}^{2}L^{2})-4\beta
L^{-2}-6n^{2}m_{p}^{2}}{3(3n^{2}m_{p}^{2}+\gamma L^{-2}\ln
(m_{p}^{2}L^{2})+\beta L^{-2})}  \nonumber \\
&&\times \left[ 1-\sqrt{\frac{3m_{p}^{2}\Omega _{\Lambda }}{%
3n^{2}m_{p}^{2}+\gamma L^{-2}\ln (m_{p}^{2}L^{2})+\beta L^{-2}}}\text{cosn}(%
\sqrt{|k|}y)\right] +\frac{b^{2}(1+\Omega _{k})}{\Omega _{\Lambda }}%
)]^{1/2}d\ln a.  \nonumber \\
&&
\end{eqnarray}%
It is interesting to note that the above potential and kinetic energy
expressions for the interacting ECHDE with the viscous generalized Chaplygin
gas coincide with the non-viscous case.

\section{Conclusion}

Enormous literature dealing with the subject of dark energy is available but
the holographic dark energy is considered to be the most promising candidate
of dark energy. In this paper we have constructed a correspondence between
the interacting ECHDE and the Chaplygin gas variants. Several candidates of
dark energy have been suggested to describe cosmic acceleration but
Chaplygin gas has emerged as a unification of dark energy and dark matter.
It's cosmic evolution is similar to initial dust like matter while it
behaves as a cosmological constant at a later epoch. In the present work, we
have investigated a model of dark energy in the presence of entropy
corrections to holographic dark energy. In this context, a link between the
ECHDE and various models of Chaplygin gas has been established. We have
found the kinetic and potential energies corresponding to each model and
also reconstructed the potentials.


\begin{thebibliography}{7}
\bibitem{1} A. G.Riess et al. [Supernova Search Team Collaboration], Astron.
J. \textbf{116} (1998)1009; S. perlmutter et al., [Supernova Cosmology
Project Collaboration], Astrophys. J. \textbf{517} (1999) 565; W. J.
Percival et al., [The 2d FGRS Collaboration], Mon .Not. Roy. Astron. Soc.
\textbf{327} (2001) 1297; P. Astier et al., Astron., Astrophs. \textbf{447}
(2006) 31; A. G. Riess et al., [Supernova Search Team Collaboration],
Astrophys. J. \textbf{607} (2004) 665; P. de Bernardis et al., Nature
(London) \textbf{404} (2000) 955; R. A Knop et al., Astrophys. J. \textbf{598%
} (2003) 102; J. L. Tonry et al., Astrophys. J. \textbf{594} (2003) 1; M. V.
Jhon Astrophys. J. \textbf{614} (2004) 1; D. N. Spergel et al., Astrophys.
J. Suppl. \textbf{170} (2007) 377; G. Hinshaw Astrophys. J. \textbf{170}
(2007) 288; M. Colless et al., Mon. Not. R. Astro. Soc. \textbf{328} (2001)
1039; M. Tegmark et al., Phys Rev. D \textbf{69} (2004) 103501; V. Springel
et al., Nature (London) \textbf{440} (2006) 1137.

\bibitem{2} P. J. E. Peebles and B. Ratra, Rev. Mod. Phys. \textbf{75}
(2003) 559; E. J. Copeland et al., Int. J. Mod. Phys. D \textbf{15} (2006)
1753; R. R. Caldwell et al., Phys. Rev. Lett. \textbf{80}, (1998) 1582; A.
R. Liddle et al., Phys. Rev. D \textbf{59} (1999) 023509; T. padmanabhan
\textbf{380} (2003) 235; A. Sen, JHEP \textbf{0207}, (2002) 065; N.N.
Weinberg et al., Phys. Rev. Lett. \textbf{91}, (2003) 071301 ; B. Feng et
al., Phys. Lett B \textbf{607} (2005) 35; Z. K. Guo et al., Phys. Lett. B
\textbf{608} (2005) 177; H. Wei et al., Class. Quantum Grav. \textbf{22}
(2005) 3189; S. M. Carrol. The Cosmological Constant, Living Rev. Rel.
\textbf{4} (2001) 1; V Sahni et al., Int. J. Mod. Phys. D \textbf{9} (2000)
373.

\bibitem{3} H. Zhang et al., Phys.Lett. B \textbf{678} (2009) 331; G.
Caldera-Cabral et al.,JCAP \textbf{07} (2009) 027; M. Jamil et al., Eur.
Phys. J. C \textbf{64} (2009) 97; M. Jamil et al., Eur. Phys. J. C \textbf{60%
} (2009) 141; M. Jamil et al., Eur. Phys. J. C \textbf{58} (2008) 111.

\bibitem{4} Sadjadi H. M. et al., Phys. Rev. D \textbf{74} (2006) 103007.

\bibitem{5} N. J. Poplawski Phys. Rev. D \textbf{74} (2006) 084032; A.
Sheykhi Phys. Lett. B \textbf{681} (2009) 205; K. Y. Kim et al., Mod. Phys.
Lett. A \textbf{22} (2007) 2631 M. R. Setare arXiv: 0909.0456; M. R. Setare
Chin. Phys. Lett. \textbf{26} (2009) 029501.

\bibitem{6} K. Enqvist, S et al., JCAP \textbf{2} (2005) 004; X. Zhang. Int.
J. Mod. Phys. D \textbf{14} (2005) 1597; D. Pavon et al., hep-th/0511053; ;
P. F. Gonzalez-Diaz, Phys. Rev. D \textbf{27} (1983) 3042.

\bibitem{7} G't Hooft, gr-qc/9310026; L. Susskind, J. Math. Phys. \textbf{%
36} (1995) 6377.

\bibitem{8} J. D. Bekenstein, Phys. Rev. D \textbf{7 }(1973) 2333; S. W. Hawking Comm. Math.
Phys. \textbf{43} (1975) 199; S. W. Hawking Phys. Rev. D \textbf{13}
(1976) 191; J. D. bekenstein, Phys. Rev. D \textbf{23 }(1981) 287;
A. G. Cohen et al., Phys. Rev. Lett. \textbf{82} (1999) 4971.

\bibitem{9} M. R. Setare Eur. Phys. J. C \textbf{50} (2007) 991; M. R.
Setare Eur. Phys. J. C \textbf{52} (2007) 689; K. Karami et al.,
arXiv: 0912.1536.

\bibitem{10} H. Wei, Commun. Theor. Phys. \textbf{52} (2009) 743; M. Jamil
and M.U. Farooq, JCAP 1003 (2010) 001.

\bibitem{11} A. Ghosh and P. Mitra, Phys. Rev. D 71 (2005) 027502; A.
Ashtekar et al, Phys. Rev. Lett. 80 (1998) 904; C. Rovelli, Phys. Rev. Lett.
\textbf{77} (1996) 3288; K. A. Miessner, Class. Quant. Grav. \textbf{21}
(2004) 5245.

\bibitem{12} L. Xu JCAP \textbf{09} (2009) 016.

\bibitem{13} H. Kim et al., Phys. Lett. B \textbf{632} (2006) 605; H. M.
Sadjadi et al., Phys. Rev. D \textbf{74} (2006) 103007.

\bibitem{14} S. D. H. Hsu, Phys. Lett. B \textbf{594} (2004) 13.

\bibitem{15} M. Li, Phys. Lett. B \textbf{603} (2004) 1.

\bibitem{16} A. Kamenshchik et al., Phys. Lett. B \textbf{511} (2001) 265.

\bibitem{17} V. Gorini et al., preprint 0403062 [gr-qc]; A. Y. Kamenshchik
et al., Phys. Lett. B \textbf{511} (2001) 265.

\bibitem{18} Sandvik et al., Phys. Rev. D \textbf{69} (2004) 123524; R. Bean
R et al., Phys. Rev. D \textbf{68} (2003) 023515.

\bibitem{19} H. B. Benaoum, preprint hep-th/0205140 (2002); ;U. Debnath et
al., Class.Qunatum Grav. \textbf{21} (2004) 5609; V. Gorini et al., Phys.
Rev. D \textbf{67} (2003) 063509; U. Alam et al., Mon. Not. Roy. Astrn. Soc.
\textbf{344} (2003) 1057; A. Dev et al., Phys Rev D \textbf{67} (2003)
023515. V. Sahni et al JETP Lett. \textbf{77} (2003) 201; M. C. Bento, et
al., Phys. Rev. D \textbf{66} (2002) 043507.

\bibitem{20} Z. K. Guo and Y. Z. Zhang, Phys. Lett. B \textbf{645} (2007)
326.

\bibitem{21} U. Debnath, Astrophys. Space Sci., \textbf{312} (2007) 295; M.
Jamil et al., Eur. J. C \textbf{61} (2009) 471.

\bibitem{22} Surajit Chattopadhyay, Ujjal Debnath, arXiv.: 0805.007v [gr-qc].

\bibitem{23} X. Zhang et al., JCAP \textbf{01} (2006) 003.

\bibitem{24} I. Brevik, S. D. Odintsov, Phys. Lett. B \textbf{455} (1999)
104; S. Nojiri, S.D. Odintsov Phys. Lett. B \textbf{562} (2003) 147.

\bibitem{25} C. Eckart, Phys. Rev. \textbf{58} (1940) 919; L. D. Landau and
E. M. Lifschitz, Fluid Mechanics (Butterworth Heineman, Oxford (1987))

\bibitem{26} C. Feng et al., Phys Lett. B \textbf{680} (2009) 355; J. Chen,
Y. Wang. :arXiv: 0904.2808v2 [gr,qc]; I. Brevik, Int. J. Mod. Phys. D
\textbf{15}, (2006) 767.

\bibitem{27} W. Zimdahl, D. Pavon, Int. J. Mod. Phys. D \textbf{15} (2006)
767.

\bibitem{28} M. Jamil, M. A. Rashid, Eur. Phys. J. C \textbf{56 }(2008) 429.
\end{thebibliography}
\end{document}